# Third harmonic generation enhancement and wavefront control using a local high-Q metasurface

Claudio U. Hail[1], Lior Michaeli[1], Harry A. Atwater[1]*

[1] Thomas J. Watson Laboratory of Applied Physics, California Institute of Technology, Pasadena, California 91125

* Correspondence and requests for materials should be addressed to H.A.A (email: haa@caltech.edu).

**Abstract**

High quality factor optical nanostructures provide great opportunity to enhance nonlinear optical processes such as third harmonic generation. However, the field enhancement in these high quality factor structures is typically accompanied by optical mode nonlocality. As a result, the enhancement of nonlinear processes comes at the cost of their local control as needed for nonlinear wavefront shaping, imaging, and holography. Here we show simultaneous strong enhancement and spatial control over third harmonic generation with a local high-Q metasurface relying on higher-order Mie-resonant modes. Our results demonstrate third harmonic generation at an efficiency of up to $3.25 \times 10^{-5}$, high quality wavefront shaping as illustrated by a third harmonic metalens, and a flatband, angle independent, third harmonic response up to $\pm 11°$ incident angle. The demonstrated high level of local control and efficient frequency conversion offer promising prospects for realizing novel nonlinear optical devices.

**Introduction**

The nonlinear optical response of materials is a key foundation for numerous light-based technologies such as light sources, optical communications, and quantum light generation[1]. In their bulk form, natural materials generally exhibit a weak optical nonlinearity per material volume, necessitating long propagation lengths to accumulate significant nonlinear interaction. Furthermore, interacting waves need to be carefully phase matched to attain high conversion efficiencies. In contrast, optical nanostructures offer a powerful route towards enhancement and control of nonlinear optical processes through diligently engineered resonant light-matter interaction, while also relaxing the phase-matching requirement. In particular, optical metasurfaces have been demonstrated as a unique platform for the miniaturized, versatile and efficient manipulation of nonlinear optical processes with subwavelength resolution[2–5]. These metasurfaces consist of a subwavelength-spaced two-dimensional array of resonant optical nanostructures that are designed to manipulate light in its various degrees of freedom. The unit cell composing the metasurface and its corresponding governing optical modes largely dictate the attainable field enhancement and nonlinear light manipulation characteristics of the surface. For example, the local surface plasmon resonance in subwavelength nanoparticles[6–8] or optically resonant dielectric nanostructures[9–12] have been employed for frequency conversion such as second harmonic (SHG) and third harmonic generation (THG). Tailoring the optical modes of these unit cells and their arrangement on the surface allows the meticulous control of amplitude, phase, polarization, and orbital angular momentum of the frequency converted light[3]. This control has enabled the realization of optical devices for nonlinear beam shaping[13], wavefront shaping[14,15], imaging[16], holography[17–19] and image encoding[20]. However, despite their excellent spatial control over nonlinear light generation, these surfaces tend to have relatively low electric field enhancement in the governing optical modes, resulting in comparatively low demonstrated conversion efficiencies between $10^{-11}$–$10^{-6}$.

Recently highly resonant optical nanostructures with high quality factor (Q) have emerged as an effective method to boost light matter interaction[21]. The high quality factors, and hence the large, attained field enhancement in these structures, make these highly suitable for amplifying nonlinear optical processes such as harmonic generation. For example, surface lattice resonances or guided mode resonances have proven effective at amplifying SHG[22–24] or THG[25]. Similarly, with the metasurface analogue of electromagnetically induced transparency (EIT) third harmonic and high harmonic generation were demonstrated[26,27]. Metasurfaces relying on quasi bound states in the continuum (q-BIC) present another intriguing approach where high quality factors can be engineered by controllably introducing assymetry or a defect into the nanostructure. Q-BIC based metasurfaces have shown enhancement of SHG[28–30], THG[31,32], high harmonic generation[33] and ultrafast nonlinear back action[34]. Recently, also the use of a multi-mode super Fano mode has been demonstrated as an alternative approach[35]. However, in these structures the gain in quality factor and thus field enhancement, is usually accompanied with mode non-locality as the mode extends over many unit cells. As a result, the increase in conversion efficiency comes at the cost of local control over light, the need for large metasurface apertures, and significant angular dispersion. A strong angular dispersion is in particular limiting, as only a narrow angular content of the excitation beam resonantly interacts with the structure, leading to a reduced field enhancement and conversion efficiency. Here we resolve this dilemma by showing that a high-Q metasurface relying on higher-order Mie resonant modes can be leveraged to attain simultaneous strong enhancement and spatial control over THG with low angular dispersion.

**Results**

Figure 1a illustrates the higher-order Mie resonant metasurface for enhancement and local control of THG. The metasurface consists of a periodic square array of amorphous silicon nanoblocks of length $L$ and height $H$ with a periodicity $P$ on a transparent borosilicate glass substrate. The nanoblocks are coated on top with a thin layer of $SiO_2$. The geometric dimensions of the structure are chosen to induce higher-order Mie resonant modes in the nanoblocks in the near infrared (NIR) spectral range, where specifically an electric dipole (ED) and electric octupole (EO) mode are dominant. Their spectral overlap creates a local high-quality factor resonance[36]. As a result of the resonant enhancement of the local electric field in the nanoblocks and the large third-order nonlinear optical susceptibility of amorphous silicon, the nonlinear optical processes are greatly amplified at resonant excitation. Illumination of the metasurface with a resonant pump at a wavelength $\lambda_R$ in the NIR spectral range leads to THG at visible wavelengths at $\lambda_R/3$. To analyze the linear optical response of the metasurface we perform finite difference time domain (FDTD) simulations. Figure 1b illustrates the simulated transmission and reflection spectrum of the metasurface with $L$ = 555 nm, $H$ = 695 nm, and $P$ = 736 nm in the NIR spectral range. A narrow dip/peak is observed in the transmission/reflection spectrum corresponding to the ED and EO modes. A quality factor of $Q$ = 189 is determined by fitting a Fano resonance lineshape, indicating a strong enhancement of the electric field in the nanostructure. To identify the dominant optical modes in the nanoblock we perform a modal decomposition of the resonant mode in the array[37]. On resonance, the electric dipole and electric octupole modes are the two largest contributions (see Supplementary Fig. 1). As in our previous work with a similar structure[36], we observe a similar low angular dispersion of less than 2.5 nm resonant wavelength shift per 10° change in incidence angle (see Supplementary Fig. 2), indicating a local nature of the resonant mode. In the spectral range of the third harmonic, the optical response of the metasurface is dominated by material absorption and no resonant modes are present (see Fig. 1c). The high quality factor, the local nature of the ED/EO mode, and its low angular dispersion make this metasurface highly suitable for enhancement and local control of THG.

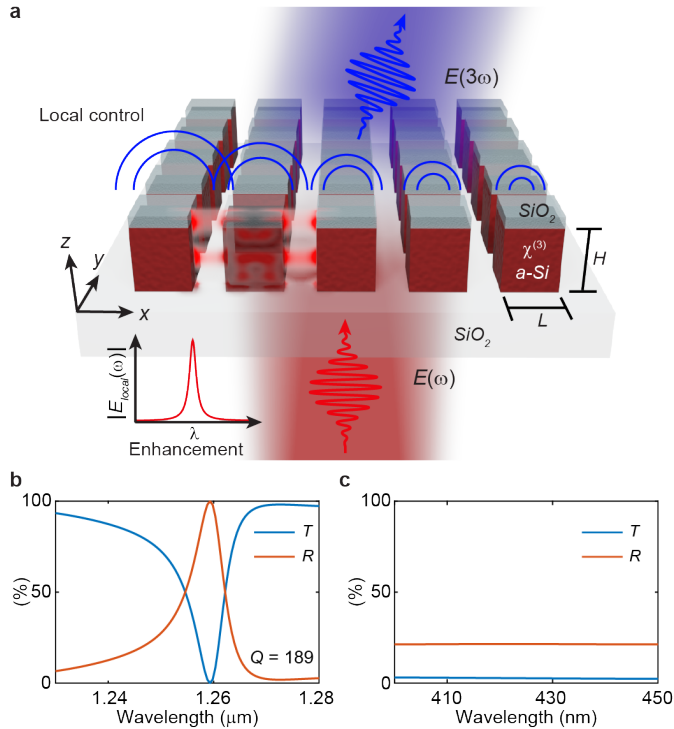

**Figure 1. | Third harmonic generation with higher-order Mie resonant metasurfaces. a,** Schematic of the higher-order Mie resonant metasurface for enhancement and local control of THG. **b,** Simulated transmission ($T$) and reflection ($R$) spectra of the metasurface with $P$ = 736 nm, $H$ = 695 nm, $L$ = 555 nm in the spectral range of the pump. **c,** Corresponding simulated transmission ($T$) and reflection ($R$) spectra of the metasurface in the spectral range of the TH.

To experimentally demonstrate the local control and enhancement of THG, we fabricate metasurfaces in a single step electron beam lithography and dry etching process on plasma deposited thin films of amorphous silicon and $SiO_2$. Figure 2a illustrates a scanning electron micrograph of a fabricated metasurface. Metasurfaces are fabricated with the nanoblock height $H$ = 695 nm, the $SiO_2$ top layer thickness is 120 nm, and the periodicity is $P$ = 736 nm. Different metasurfaces with nanoblock side lengths of $L$ = [548, 561, 572, 589, 604, 619, 639] nm are fabricated, each with a size of 150 μm × 150 μm. We perform a linear optical characterization of the metasurfaces in the NIR spectral range by illuminating the structures from the substrate side with a loosely focused beam from a white light supercontinuum source. The transmitted light is collected with a microscope objective lens (50x, 0.95 NA) and projected onto a NIR grating spectrometer. For the nonlinear optical characterization, we pump the surface with a 100-fs laser pulse (10 kHz repetition rate) at varying wavelengths in the NIR spectral range using the same illumination and collection optics. We filter the pump from the collected light with a set of short-pass filters and project the remaining THG on to a visible light grating spectrometer. By imaging the sample plane on an InGaAs camera we monitor the metasurface position and the position, size and angular content of the pump beam. A detailed description of the experimental set-up is given in the Methods section.

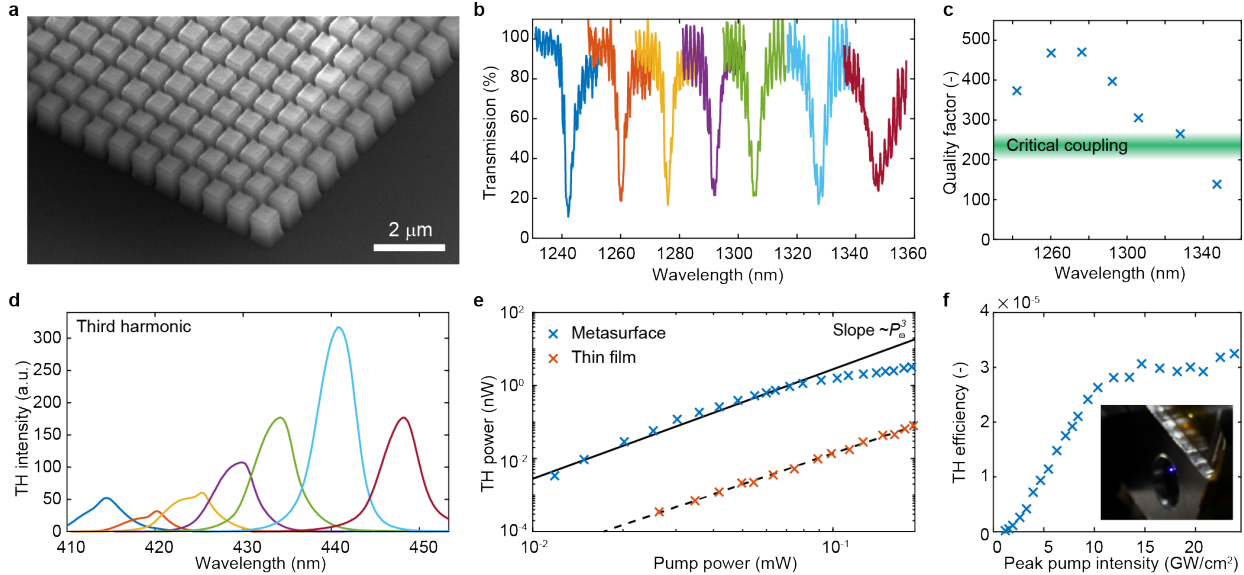

**Figure 2. | Linear and nonlinear optical characterization of the high-Q metasurfaces. a,** Scanning electron micrograph of a metasurface. **b,** Experimentally measured linear transmission spectra of higher-order Mie resonant metasurfaces with varying nanoblock side length $L$ = [548, 561, 572, 589, 604, 619, 639] nm from left to right. For all metasurfaces the dimensions are $P$ = 736 nm and $H$ = 695 nm. **c,** Quality factors of the metasurfaces in (**b**) as determined from a Fano fit. **d,** Experimentally measured third-harmonic spectra of the corresponding metasurfaces in (**b**) in corresponding line color. **e,** Experimentally measured third harmonic power as a function of incident pump power for the critically coupled metasurface ($L$ = 619 nm), and a reference unpatterned layer stack with identical geometrical parameters. The solid line shows the ideal line of a third order nonlinear process with the $P_\omega^3$ power dependence. Beyond 80 μW, a typical saturation behavior is observed. The dashed line shows a least-squares fit to the thin film data resulting a power dependence of $P_\omega^{3.04}$. **f,** Measured efficiency of THG vs pump intensity. The inset shows a photograph of the metasurface, with the third harmonic light clearly visible to the naked eye.

Figure 2b shows the results of the linear optical characterization of the metasurfaces, illustrating the measured transmission spectra of the metasurfaces with varying nanoblock lengths. Narrow spectral resonances are observed in transmission spanning a spectral range of 1240 nm to 1360 nm when varying the nanoblock side length in the range of $L$ = 548–639 nm. The high frequency oscillations in the measured transmission are due to the interference of the light in the thin glass substrate of the metasurface. The quality factors of the corresponding resonances are shown in Fig. 2c and range between 139–470, as determined by fitting a Fano lineshape to the transmission spectra[38]. This indicates that the metasurface produces a significant electric field enhancement, making it a favorable choice for THG amplification.

Figure 2d illustrates the measured spectra of the THG of the corresponding metasurfaces in Fig. 2b. All metasurfaces are resonantly excited with a constant peak pump intensity of 9.5 GW/cm$^2$. The corresponding spectrum of the pump beam for each metasurface is illustrated in Supplementary Fig. 3. A strong THG signal centered around one third of the resonant wavelength is recorded for each metasurface. Notably, the intensity of the THG varies considerably between the different curves, with the largest signal recorded for a nanoblock side length $L$ = 619 nm. This is attributed to the critical coupling condition, where the quality factor is approximately one half of the maximum quality factor[31]. Therefore, adjusting geometrical parameters, such as the nanoblock side length, allows for optimizing the THG of the metasurface by ensuring that it operates at the critical coupling condition.

As a next step, we consider the critically coupled metasurface ($L$ = 619 nm), which shows the strongest TH signal according to Fig. 2d and vary the pump power, $P_\omega$, while recording the collected TH power, $P_{3\omega}$. Figure 2e shows the measured TH power for varying pump power of this metasurface. At pump powers below 80 μW, the THG power approximately follows a power of 3 trend, as expected from a third-order nonlinear process. At powers above 80 μW, the THG power saturates and follows a reduced power trend. This saturation behavior is typically observed in silicon-based metasurfaces[31] and is attributed to two-photon and free-carrier absorption in the amorphous silicon. For comparison, we measured the power dependence of THG from a reference unpatterned thin film of amorphous silicon and $SiO_2$ with the same thickness as in the metasurface. As illustrated in Fig. 2e, a similar trend with pump power is obtained but at significantly lower TH power levels. The comparison shows that the metasurface enhances THG by up to a factor of 240, even though the volume of amorphous silicon is less as compared to the thin film. From the measurement we determine the THG efficiency η = $P_{3\omega}/P_\omega$. Figure 2f illustrates the measured THG efficiency for varying peak pump intensities. Similar to the THG power, the efficiency increases at low peak pump intensities until it saturates above 10 GW/cm$^2$. At 10.3 GW/cm$^2$ we record an efficiency of 2.65 × 10$^{-5}$, and a maximum efficiency of 3.25 × 10$^{-5}$ is recorded at 23.8 GW/cm$^2$. This is more than an order of magnitude higher than efficiencies recorded in previous works. For example, an efficiency of 1.2 × 10$^{-6}$ at pump intensities of 3.2 GW/cm$^2$ was reported with metasurface using the analogue of EIT[26]. In another recent study, an efficiency of 2.8 × 10$^{-7}$ at peak pump intensities of 1.2 GW/cm$^2$ was reported[35]. It is worth noting that higher pump intensities are used here, however, in previous works TH power generally saturated at lower peak power intensity as compared to our reporting. At similar peak pump intensities of 3.3 GW/cm$^2$, we measure an efficiency of 4.2 × 10$^{-6}$. The effective THG efficiency is likely higher than the measured one, since in experiment only THG emitted to the air side is collected. Numerical simulations suggest that only about 57% of the total THG is emitted to the air half space and collected by the objective lens. Additionally, exciting with a fs pulse that is spectrally matched to the metasurface resonance, or red-shifting the resonance to a spectral region of lower optical loss at the TH wavelength is also expected to yield higher efficiencies. Remarkably, the TH intensity is at a level where it can be observed by naked eye, as the blue light captured in a photograph of the metasurface shows in the inset of Fig. 2f.

The independence of THG enhancement from the pump illumination angle is a desirable characteristic, as it enables attaining considerable TH powers at low average pump powers by focusing the pump beam. Namely, illumination with a wide range of angles decreases the illumination spot size, resulting in an increased pump intensity and hence TH intensity. To gain insight into the TH power as function of the angular content of the incident beam, we analytically calculate the TH power emitted from a thin (<<λ) metasurface for the ideal case of an angle-independent effective third-order nonlinear susceptibility, $\chi^{(3)}_{\text{eff}}$. Assuming the surface is illuminated by a gaussian pump beam, we obtain the TH power by integration over the area of the gaussian beam in the focal plane[1,39]

$$P_{3\omega} = \int \frac{9\varepsilon_0 \omega^2 d^2}{128 n_{3\omega} c} \left| \chi^{(3)}_{eff} E^3_\omega(r) \right|^2 dA, \qquad (1)$$

where $E_\omega$ represents the electric field of the pump, $c$ the speed of light, $d$ the thickness of the metasurface, $n_{3\omega}$ the refractive index at the TH wavelength, and $\varepsilon_0$ the vacuum permittivity. By incorporating a gaussian electric field profile with beam width $w$ and a power $P_\omega$, the integral is evaluated to obtain

$$P_{3\omega} = \frac{3\omega^2 d^2 P_\omega^3}{4 n_{3\omega} n_\omega^3 \varepsilon_0^2 c^4 \pi^2 w^4} \left| \chi^{(3)}_{eff} \right|^2 \sim NA^4, \qquad (2)$$

where $n_\omega$ is the refractive index at the pump wavelength, and $NA = \sin\theta$, is the numerical aperture of the pump illumination with a half angle $\theta$ of the incident light cone. A more elaborate derivation is provided in Supplementary Note 1. As evident from Eq. 2, at a constant pump power, the TH power increases with the fourth power of the numerical aperture of the pump focusing. Maximizing TH power therefore requires maximizing the numerical aperture. However, in optical metasurfaces the enhancement of THG typically relies on extended high quality factor modes, such as guided modes[25], surface lattice modes[22], EIT-analogue[26], or symmetry-protected q-BIC[31,32], that are angle sensitive due to a large angular dispersion. The local ED/EO mode employed in our structure shows a flat angular dispersion of less than 2.5 nm resonance wavelength shift per 10° change in incidence angle (see Supplementary Fig. 2), and hence is particularly suitable for THG with focused illumination.

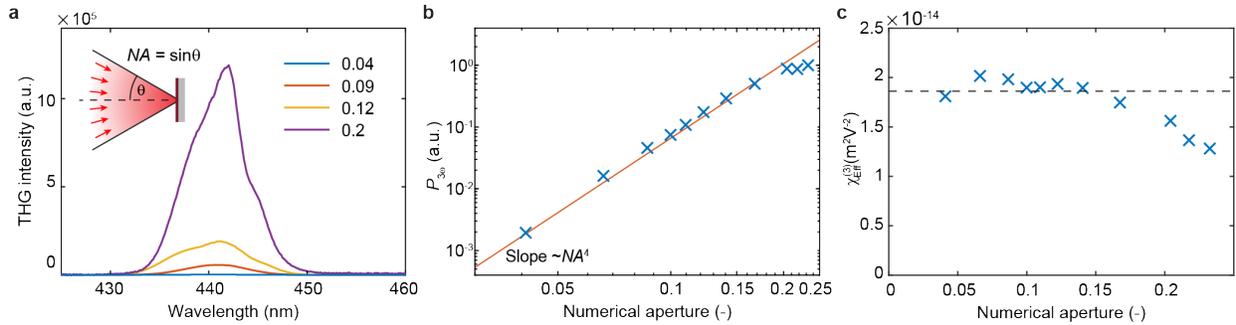

**Figure 3. | Angle-independent third harmonic generation. a,** Experimentally measured third-harmonic spectrum for varying numerical aperture of the illuminating pump beam scaled to a constant pump power of 38 µW. **b,** Experimentally measured power of the third harmonic with varying numerical aperture of the illuminating pump beam. The solid curve with a slope $NA^4$ represents the power dependence for metasurface with an angle-independent $\chi^{(3)}_{eff}$. **c,** Calculated effective third-order nonlinear optical susceptibility of the metasurface as a function of the numerical aperture of the illuminating pump beam.

Figure 3 shows the experimental results of the pump-angle dependent analysis of THG from the metasurface with $L$ = 619 nm at resonant excitation. In the experiment, we illuminate the metasurface with a focused pump beam and set the numerical aperture of the illumination in the range of 0.04–0.25 with an iris that defines the aperture of illumination into the objective. The pump power and TH power are recorded for each numerical aperture value, and the collected TH signal is the scaled by $P_{\omega,0}^3 / P_\omega^3$ to the same incident pump power $P_{\omega,0}$. For all measurements the peak pump intensity is maintained well below the saturation value observed in Fig. 2e. The measured THG spectra for different numerical apertures scaled to a fixed pump power are illustrated in Fig. 3a. A large increase in TH signal is observed when increasing the numerical aperture from 0.04 to 0.2. Figure 3b shows the measured TH power for varying numerical aperture of the pump beam. The measured TH power increases approximately with $\sim NA^4$ up to a value of $NA = 0.2$. This suggests that the TH power indeed follows Eq. 2 and indicates that $\chi^{(3)}_{eff}$ shows a flatband dispersion with respect to the incident angle below NA < 0.2. To further confirm this behavior, we calculate an effective third-order susceptibility, $\chi^{(3)}_{eff}(NA)$, as a function of the numerical aperture from our measurement results. This is done by inserting the measured pump beam intensity profile into Eq. 1 and solving for $\chi^{(3)}_{eff}(NA)$ (see Supplementary Note 2). Figure 3c shows the calculated effective susceptibility for varying NA. A value of $\chi^{(3)}_{eff} \sim 1.86 \times 10^{-14}$ m$^2$/V$^2$ is determined, which is 75,918 times larger than $\chi^{(3)}$ of amorphous silicon. The extracted $\chi^{(3)}_{eff}(NA)$ remains approximately constant with NA up to a value of $NA = 0.2$, corresponding to a cone angle of $\theta = 11.5°$, beyond which a slow decrease with NA sets in.

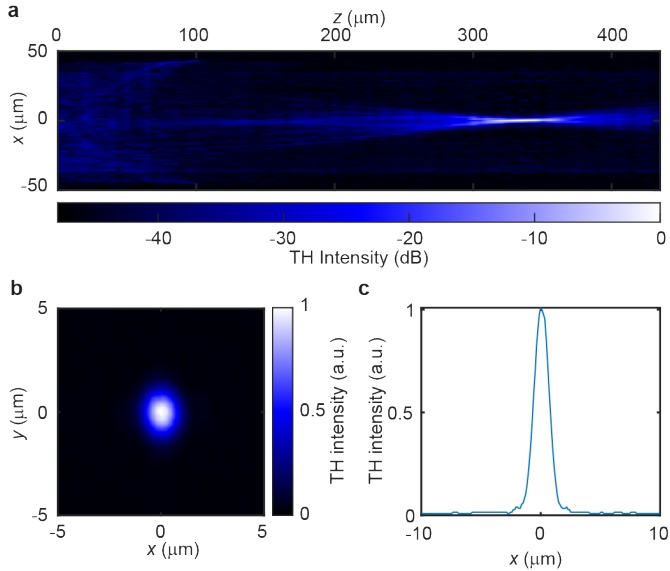

**Figure 4. | Enhanced third harmonic wavefront manipulation. a,** Measured field intensity of the third harmonic at $\lambda = 442$ nm along the optical axis in the x-z plane of a metalens. **b,** Measured field intensity in the focal plane (x-y plane) of the third harmonic of the metalens in (**a**). **c,** Cross section of the measured field intensity in the focal plane in (**b**). The metalens is 100 μm in diameter and $P = 736$ nm and $H = 695$ nm.

The local ED/EO mode employed for THG enhancement enables local control over the TH wavefront by tailoring the unit cell dimensions along the interface. This is done by controlling the phase of the emitted TH by setting the nanoblock side length. The specific TH phase depends on the ED/EO mode, the material dispersion, and the optical response of the structure at the TH wavelength. We model these effects using a finite element numerical simulation and determine the TH phase for a specific nanoblock side length to create a look up table of $\varphi_{TH}$ vs. $L$ (see Fig. S4). To demonstrate the TH wavefront shaping capabilities we realize a TH metalens by imposing a paraboloidal phase profile on the TH emitted to the air half space of the metasurface. Figure 4a illustrates the measured TH field intensity in a cross section along the optical axis. Figure 4b and c show the TH field intensity in the focal plane and a line cross section through the focus. A symmetric focusing of the TH with high quality and negligible stray light in the focal plane is observed. As opposed to previous demonstrations of TH wavefront shaping[13–18], the structure here allows for simultaneous enhancement of THG with a high-Q mode and local control for wavefront manipulation.

In summary, we have demonstrated how a local high-Q mode can be leveraged to strongly enhance THG, and simultaneously offer local control over THG, as exemplified with TH wavefront shaping and flatband dispersion nonlinearity. Our results showcase a remarkable enhancement of THG as evidenced by a maximum conversion efficiency of $3 \times 10^{-5}$. Due to the local response of our metasurface we obtain an approximately angle independent THG up to 11° and high-quality wavefront control, as demonstrated with a TH metalens. Compared to the current state of the art of metasurfaces for THG[40] we show the highest reported TH conversion efficiencies to date and additionally enable wavefront manipulation of THG (see Supplementary Table 1 for a detailed comparison). The high TH conversion efficiency, large degree of local TH control, and angle independent performance make our approach ideal for applications in optical signal processing, nonlinear imaging, and data storage. Furthermore, integrating active reconfiguration into the metasurface may allow for dynamic steering of THG at up to microsecond timescales or faster[41–43]. Besides THG, we envision the high-Q metasurface presented here to have a profound impact on broader areas of nonlinear optics such as high-harmonic generation, all-optical switching, and down-conversion processes.

## Methods

**Experiment** The fabricated metasurfaces were characterized on a home-built optical transmission microscope (schematically illustrated in Supplementary Fig. 5). For the linear optical characterization, coherent light from a supercontinuum laser (NKT, Super K Extreme) was loosely focused on the metasurface with a lens of $f$ = 50 mm. The transmitted light was collected with an objective lens (50x, 0.95 NA, Zeiss) and projected on to a NIR grating spectrometer (Princeton Instruments, Acton 2500, PylonNIR). For the nonlinear optical characterization, a Ti:Sapphire laser with 100 fs pulse length and 10 kHz repetition rate (Coherent Libra) was used to drive an optical parametric amplifier (Coherent, OPerA Solo) to generate a NIR pump beam within the wavelength range 1.2–1.6 $\mu$m. The pump beam was loosely focused on the metasurface with a lens of $f$ = 50 mm, the transmitted light was filtered with two short pass filters (Thorlabs FGS900), and the remaining TH beam was projected onto a visible grating spectrometer (Princeton Instruments, Acton 2500, Pixis). The pump spot size was 100 $\mu$m unless noted otherwise. The pump power was measured by placing a power meter (Thorlabs S122C) at the location of the metasurface. The third harmonic power was measured by integrating the intensity vs wavelength obtained by the grating spectrometer camera. For this the spectrometer counts were converted to optical power by performing a calibration with a power meter (Thorlabs S130C). For the metalens characterization, the focal plane was imaged on a CMOS camera (Thorlabs DCC1645C-HQ) and a scan along the optical axis was performed by moving the surface along the $z$ direction. The pump intensity and beam profile were measured by projecting the collected pump beam on to an InGaAs camera without the low pass filters in place. For the experiments reported in Fig. 3 an objective lens (10x, 0.25 NA, Olympus) was used to focus the pump on the sample. With an overfilled aperture, the NA was set with an iris placed directly before the objective lens.

Scanning electron micrographs were acquired on an FEI Nova 600 NanoLab system to measure the sizes of the fabricated structures. For imaging, the surfaces were covered with a 4 nm thick carbon layer by sputter deposition.

**Fabrication** The metasurfaces were fabricated on borosilicate glass substrates ($n$ = 1.503) with a thickness of 220 µm. To remove organic residues from the surface, the substrates were cleaned in an ultrasonic bath in acetone, isopropyl alcohol, and deionized water each for 15 min, dried using a $N_2$ gun. Amorphous silicon was deposited onto the glass using plasma-enhanced chemical vapor deposition. In a subsequent step, the nanoblocks were written in a spin coated MaN-2403 resist layer by standard electron beam lithography. The nanoblocks were then transferred to the amorphous silicon using an $SiO_2$ hard mask with chlorine-based inductively coupled reactive ion etching. The uniform metasurfaces were fabricated on an area of 150 µm × 150 µm. The metalens was fabricated with a diameter of 100 µm and a parabolic phase profile according to the equation

$$\varphi(x,y) = \frac{2\pi}{\lambda}\left(\sqrt{x^2 + y^2 + f^2} - f\right), \tag{1}$$

where $\lambda$ is the design wavelength ($\lambda$ = 1327 nm) and $f$ the focal length. The variation of the nanoblock side length is set with a discretization of 0.1 nm according to Supplementary Fig. 4. Further metalens design parameters are given in Supplementary Table 2.

**Simulation** The numerical modelling of the nanostructures was carried out using an FDTD method. Simulations were performed with a commercially available FDTD software (Lumerical FDTD Solutions). In the NIR spectral range a constant refractive index of $n$ = 1.503 was used for the borosilicate glass and $n$ = 1.453 for the $SiO_2$. In the TH spectral range $n$ = 1.53 and $n$ = 1.47 are used, respectively. For amorphous silicon experimentally measured values were used as determined by ellipsometry (see Supplementary Fig. 6). The simulations were carried out with a spatially coherent plane wave illumination, a perfectly matched layer boundary condition along the $z$ direction, and periodic boundary conditions were applied along the $x$ and $y$ direction. For the nanoblock a smallest mesh-refinement of 5 nm was used.

The third harmonic phase was determined from finite element simulations by solving Maxwell's equations in the frequency domain using the commercially available software COMSOL Multiphysics. A first simulation is performed over the spectral range of the pump to obtain the electric field in the silicon nanoblocks, and thereby the nonlinear polarization $\boldsymbol{P}_{NL}(3\omega) = \varepsilon_0\chi^{(3)}\boldsymbol{E}^3(\omega)$. A second simulation is then performed in the TH spectral range introducing a current source generated by $\boldsymbol{P}_{NL}$. For the simulations we adopt the undepleted pump approximation[1], we approximate the susceptibility tensor for amorphous silicon to a scalar non-dispersive value of $\chi^{(3)}$ = 2.45 × 10$^{-19}$ m$^2$/V$^2$, [15,31] and we use periodic boundary conditions along the $x$ and $y$ direction, and a perfectly matched layer along the $z$ direction.

**References**

(1) Boyd, R. W. *Nonlinear Optics*; Academic Press: Burlington, MA, 2008.
(2) Smirnova, D.; Kivshar, Y. S. Multipolar Nonlinear Nanophotonics. *Optica* **2016**, *3* (11), 1241. https://doi.org/10.1364/optica.3.001241.
(3) Keren-Zur, S.; Avayu, O.; Michaeli, L.; Ellenbogen, T. Shaping Light with Nonlinear Metasurfaces. *Adv. Opt. Photonics* **2016**, *10* (1), 309–353. https://doi.org/10.1364/aop.10.000309.
(4) Li, G.; Zhang, S.; Zentgraf, T. Nonlinear Photonic Metasurfaces. *Nat. Rev. Mater.* **2017**, *2*

(5), 1–14. https://doi.org/10.1038/natrevmats.2017.10.
(5) Krasnok, A.; Tymchenko, M.; Alù, A. Nonlinear Metasurfaces: A Paradigm Shift in Nonlinear Optics. *Mater. Today* **2018**, *21* (1), 8–21. https://doi.org/10.1016/j.mattod.2017.06.007.
(6) Thyagarajan, K.; Rivier, S.; Lovera, A.; Martin, O. J. F. Enhanced Second-Harmonic Generation from Double Resonant Plasmonic Antennae. *Opt. Express* **2012**, *20* (12), 12860. https://doi.org/10.1364/oe.20.012860.
(7) Thyagarajan, K.; Butet, J.; Martin, O. J. F. Augmenting Second Harmonic Generation Using Fano Resonances in Plasmonic Systems. *Nano Lett.* **2013**, *13* (4), 1847–1851. https://doi.org/10.1021/nl400636z.
(8) Zhang, Y.; Grady, N. K.; Ayala-Orozco, C.; Halas, N. J. Three-Dimensional Nanostructures as Highly Efficient Generators of Second Harmonic Light. *Nano Lett.* **2011**, *11* (12), 5519–5523. https://doi.org/10.1021/nl2033602.
(9) Shcherbakov, M. R.; Neshev, D. N.; Hopkins, B.; Shorokhov, A. S.; Staude, I.; Melik-Gaykazyan, E. V.; Decker, M.; Ezhov, A. A.; Miroshnichenko, A. E.; Brener, I.; Fedyanin, A. A.; Kivshar, Y. S. Enhanced Third-Harmonic Generation in Silicon Nanoparticles Driven by Magnetic Response. *Nano Lett.* **2014**, *14* (11), 6488–6492. https://doi.org/10.1021/nl503029j.
(10) Shorokhov, A. S.; Melik-Gaykazyan, E. V.; Smirnova, D. A.; Hopkins, B.; Chong, K. E.; Choi, D. Y.; Shcherbakov, M. R.; Miroshnichenko, A. E.; Neshev, D. N.; Fedyanin, A. A.; Kivshar, Y. S. Multifold Enhancement of Third-Harmonic Generation in Dielectric Nanoparticles Driven by Magnetic Fano Resonances. *Nano Lett.* **2016**, *16* (8), 4857–4861. https://doi.org/10.1021/acs.nanolett.6b01249.
(11) Grinblat, G.; Li, Y.; Nielsen, M. P.; Oulton, R. F.; Maier, S. A. Enhanced Third Harmonic Generation in Single Germanium Nanodisks Excited at the Anapole Mode. *Nano Lett.* **2016**, *16* (7), 4635–4640. https://doi.org/10.1021/acs.nanolett.6b01958.
(12) Grinblat, G.; Li, Y.; Nielsen, M. P.; Oulton, R. F.; Maier, S. A. Efficient Third Harmonic Generation and Nonlinear Subwavelength Imaging at a Higher-Order Anapole Mode in a Single Germanium Nanodisk. *ACS Nano* **2017**, *11* (1), 953–960. https://doi.org/10.1021/acsnano.6b07568.
(13) Keren-Zur, S.; Avayu, O.; Michaeli, L.; Ellenbogen, T. Nonlinear Beam Shaping with Plasmonic Metasurfaces. *ACS Photonics* **2016**, *3* (1), 117–123. https://doi.org/10.1021/acsphotonics.5b00528.
(14) Li, G.; Chen, S.; Pholchai, N.; Reineke, B.; Wong, P. W. H.; Pun, E. Y. B.; Cheah, K. W.; Zentgraf, T.; Zhang, S. Continuous Control of the Nonlinearity Phase for Harmonic Generations. *Nat. Mater.* **2015**, *14* (6), 607–612. https://doi.org/10.1038/nmat4267.
(15) Wang, L.; Kruk, S.; Koshelev, K.; Kravchenko, I.; Luther-Davies, B.; Kivshar, Y. Nonlinear Wavefront Control with All-Dielectric Metasurfaces. *Nano Lett.* **2018**, *18* (6), 3978–3984. https://doi.org/10.1021/acs.nanolett.8b01460.
(16) Schlickriede, C.; Kruk, S. S.; Kruk, S. S.; Wang, L.; Sain, B.; Kivshar, Y.; Zentgraf, T. Nonlinear Imaging with All-Dielectric Metasurfaces. *Nano Lett.* **2020**, *20* (6), 4370–4376. https://doi.org/10.1021/acs.nanolett.0c01105.
(17) Ye, W.; Zeuner, F.; Li, X.; Reineke, B.; He, S.; Qiu, C. W.; Liu, J.; Wang, Y.; Zhang, S.; Zentgraf, T. Spin and Wavelength Multiplexed Nonlinear Metasurface Holography. *Nat. Commun.* **2016**, *7* (May), 1–7. https://doi.org/10.1038/ncomms11930.
(18) Gao, Y.; Fan, Y.; Wang, Y.; Yang, W.; Song, Q.; Xiao, S. Nonlinear Holographic All-

Dielectric Metasurfaces. *Nano Lett.* **2018**, *18* (12), 8054–8061. https://doi.org/10.1021/acs.nanolett.8b04311.

(19) Reineke, B.; Sain, B.; Zhao, R.; Carletti, L.; Liu, B.; Huang, L.; De Angelis, C.; Zentgraf, T. Silicon Metasurfaces for Third Harmonic Geometric Phase Manipulation and Multiplexed Holography. *Nano Lett.* **2019**, *19* (9), 6585–6591. https://doi.org/10.1021/acs.nanolett.9b02844.

(20) Walter, F.; Li, G.; Meier, C.; Zhang, S.; Zentgraf, T. Ultrathin Nonlinear Metasurface for Optical Image Encoding. *Nano Lett.* **2017**, *17* (5), 3171–3175. https://doi.org/10.1021/acs.nanolett.7b00676.

(21) Barton, D.; Hu, J.; Dixon, J.; Klopfer, E.; Dagli, S.; Lawrence, M.; Dionne, J. High-Q Nanophotonics: Sculpting Wavefronts with Slow Light. *Nanophotonics* **2020**, *10* (1), 83–88. https://doi.org/10.1515/nanoph-2020-0510.

(22) Michaeli, L.; Keren-Zur, S.; Avayu, O.; Suchowski, H.; Ellenbogen, T. Nonlinear Surface Lattice Resonance in Plasmonic Nanoparticle Arrays. *Phys. Rev. Lett.* **2017**, *118* (24), 1–6. https://doi.org/10.1103/PhysRevLett.118.243904.

(23) Abir, T.; Tal, M.; Ellenbogen, T. Second-Harmonic Enhancement from a Nonlinear Plasmonic Metasurface Coupled to an Optical Waveguide. *Nano Lett.* **2022**, *22* (7), 2712–2717. https://doi.org/10.1021/acs.nanolett.1c04584.

(24) Qu, L.; Bai, L.; Jin, C.; Liu, Q.; Wu, W.; Gao, B.; Li, J.; Cai, W.; Ren, M.; Xu, J. Giant Second Harmonic Generation from Membrane Metasurfaces. *Nano Lett.* **2022**, *22* (23), 9652–9657. https://doi.org/10.1021/acs.nanolett.2c03811.

(25) Chen, S.; Rahmani, M.; Li, K. F.; Miroshnichenko, A.; Zentgraf, T.; Li, G.; Neshev, D.; Zhang, S. Third Harmonic Generation Enhanced by Multipolar Interference in Complementary Silicon Metasurfaces. *ACS Photonics* **2018**, *5* (5), 1671–1675. https://doi.org/10.1021/acsphotonics.7b01423.

(26) Yang, Y.; Wang, W.; Boulesbaa, A.; Kravchenko, I. I.; Briggs, D. P.; Puretzky, A.; Geohegan, D.; Valentine, J. Nonlinear Fano-Resonant Dielectric Metasurfaces. *Nano Lett.* **2015**, *15* (11), 7388–7393. https://doi.org/10.1021/acs.nanolett.5b02802.

(27) Liu, H.; Guo, C.; Vampa, G.; Zhang, J. L.; Sarmiento, T.; Xiao, M.; Bucksbaum, P. H.; Vučković, J.; Fan, S.; Reis, D. A. Enhanced High-Harmonic Generation from an All-Dielectric Metasurface. *Nat. Phys.* **2018**, *14* (10), 1006–1010. https://doi.org/10.1038/s41567-018-0233-6.

(28) Vabishchevich, P. P.; Liu, S.; Sinclair, M. B.; Keeler, G. A.; Peake, G. M.; Brener, I. Enhanced Second-Harmonic Generation Using Broken Symmetry III-V Semiconductor Fano Metasurfaces. *ACS Photonics* **2018**, *5* (5), 1685–1690. https://doi.org/10.1021/acsphotonics.7b01478.

(29) Liu, Z.; Xu, Y.; Lin, Y.; Xiang, J.; Feng, T.; Cao, Q.; Li, J.; Lan, S.; Liu, J. High- Q Quasibound States in the Continuum for Nonlinear Metasurfaces. *Phys. Rev. Lett.* **2019**, *123* (25), 1–6. https://doi.org/10.1103/PhysRevLett.123.253901.

(30) Han, Z.; Ding, F.; Cai, Y.; Levy, U. Significantly Enhanced Second-Harmonic Generations with All-Dielectric Antenna Array Working in the Quasi-Bound States in the Continuum and Excited by Linearly Polarized Plane Waves. *Nanophotonics* **2021**, *10* (3), 1189–1196.

(31) Koshelev, K.; Tang, Y.; Li, K.; Choi, D. Y.; Li, G.; Kivshar, Y. Nonlinear Metasurfaces Governed by Bound States in the Continuum. *ACS Photonics* **2019**, *6* (7), 1639–1644. https://doi.org/10.1021/acsphotonics.9b00700.

(32) Yang, G.; Dev, S. U.; Allen, M. S.; Allen, J. W.; Harutyunyan, H. Optical Bound States in the Continuum Enabled by Magnetic Resonances Coupled to a Mirror. *Nano Lett.* **2022**, *22* (5), 2001–2008. https://doi.org/10.1021/acs.nanolett.1c04764.

(33) Zograf, G.; Koshelev, K.; Zalogina, A.; Korolev, V.; Hollinger, R.; Choi, D. Y.; Zuerch, M.; Spielmann, C.; Luther-Davies, B.; Kartashov, D.; Makarov, S. V.; Kruk, S. S.; Kivshar, Y. High-Harmonic Generation from Resonant Dielectric Metasurfaces Empowered by Bound States in the Continuum. *ACS Photonics* **2022**, *9* (2), 567–574. https://doi.org/10.1021/acsphotonics.1c01511.

(34) Sinev, I. S.; Koshelev, K.; Liu, Z.; Rudenko, A.; Ladutenko, K.; Shcherbakov, A.; Sadrieva, Z.; Baranov, M.; Itina, T.; Liu, J.; Bogdanov, A. A.; Kivshar, Y. Observation of Ultrafast Self-Action Effects in Quasi-BIC Resonant Metasurfaces. *Nano Lett.* **2021**, *21* (20), 8848–8855. https://doi.org/10.1021/acs.nanolett.1c03257.

(35) Hähnel, D.; Golla, C.; Albert, M.; Zentgraf, T.; Myroshnychenko, V.; Förstner, J.; Meier, C. A Multi-Mode Super-Fano Mechanism for Enhanced Third Harmonic Generation in Silicon Metasurfaces. *Light Sci. Appl.* **2023**, *12* (1), 23–26. https://doi.org/10.1038/s41377-023-01134-1.

(36) Hail, C. U.; Foley, M.; Sokhoyan, R.; Atwater, H. A. High Quality Factor Metasurfaces for Two Dimensional Wavefront Manipulation. arXiv Preprint at https://arxiv.org/abs/2212.05647. **2023**.

(37) Savinov, V.; Fedotov, V. A.; Zheludev, N. I. Toroidal Dipolar Excitation and Macroscopic Electromagnetic Properties of Metamaterials. *Phys. Rev. B - Condens. Matter Mater. Phys.* **2014**, *89* (20). https://doi.org/10.1103/PhysRevB.89.205112.

(38) Bogdanov, A. A.; Koshelev, K. L.; Kapitanova, P. V.; Rybin, M. V.; Gladyshev, S. A.; Sadrieva, Z. F.; Samusev, K. B.; Kivshar, Y. S.; Limonov, M. F. Bound States in the Continuum and Fano Resonances in the Strong Mode Coupling Regime. *Adv. Photonics* **2019**, *1* (01), 1. https://doi.org/10.1117/1.ap.1.1.016001.

(39) Kumar, N.; Kumar, J.; Gerstenkorn, C.; Wang, R.; Chiu, H.; Smirl, A. L.; Zhao, H. Third Harmonic Generation in Graphene and Few-Layer Graphite Films. *Phys. Rev. B* **2013**, *87*, 121406. https://doi.org/10.1103/PhysRevB.87.121406.

(40) Vabishchevich, P.; Kivshar, Y. Nonlinear Photonics with Metasurfaces. *Photonics Res.* **2023**, *11* (2), B50. https://doi.org/10.1364/prj.474387.

(41) Yu, J.; Park, S.; Hwang, I.; Kim, D.; Demmerle, F.; Boehm, G.; Amann, M. C.; Belkin, M. A.; Lee, J. Electrically Tunable Nonlinear Polaritonic Metasurface. *Nat. Photonics* **2022**, *16* (1), 72–78. https://doi.org/10.1038/s41566-021-00923-7.

(42) Sokhoyan, R.; Hail, C. U.; Foley, M.; Grajower, M.; Atwater, H. A. All-Dielectric High-Q Dynamically Tunable Transmissive Metasurfaces. arXiv preprint at https://arxiv.org/abs/2309.08031 **2023**.

(43) Sharma, M.; Tal, M.; McDonnell, C.; Ellenbogen, T. Electrically and All-Optically Switchable Nonlocal Nonlinear Metasurfaces. *Sci. Adv.* **2023**, *9* (33), eadh2353. https://doi.org/10.1126/sciadv.adh2353.

**Acknowledgements**

This work was supported by the Air Force Office of Scientific Research under grant FA9550-18-1-0354 and the Meta-Imaging MURI grant #FA9550-21-1-0312. C.U.H. also acknowledges support from the Swiss National Science Foundation through the Early Postdoc Mobility Fellowship grant #P2EZP2_191880 and the Postdoc Mobility grant #P500PT_214452. L.M.

acknowledges support from the Fulbright Fellowship program and the Breakthrough Foundation. We gratefully acknowledge the critical support and infrastructure provided for this work by The Kavli Nanoscience Institute at Caltech.

**Contributions**

C.U.H, L.M. and H.A.A. conceived the project. C.U.H preformed the simulations, fabricated the devices, built the experiment, performed the measurements, and analyzed the results. L.M. assisted in experiment design, simulations, and data analysis and interpretation. C.U.H. wrote the manuscript with input from all other authors. H.A.A supervised all aspects of the project.

**Competing interests**

The authors declare no competing interests.

**Correspondence and requests for materials** should be addressed to H.A.A.

# Supplementary Information: Third harmonic generation enhancement and wavefront control using a local high-Q metasurface

Claudio U. Hail[1], Lior Michaeli[1], Harry A. Atwater[1]*

[1] Thomas J. Watson Laboratory of Applied Physics, California Institute of Technology, Pasadena, California 91125

* Correspondence and requests for materials should be addressed to H.A.A (email: haa@caltech.edu).

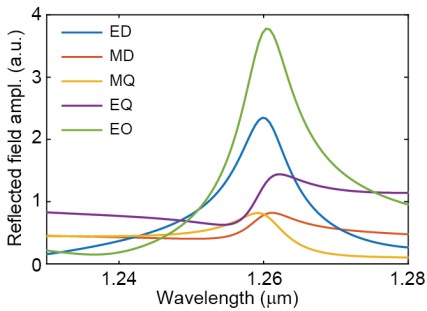

**Supplementary Figure 1 | Multipole expansion of the reflected field amplitude.** Calculated contributions of the electric dipole (ED), magnetic dipole (MD), magnetic quadrupole (MQ), electric quadrupole (EQ), and electric octupole (EO) to the reflected field amplitude determined from a multipole expansion of a nanoblock within a periodic array[1]. The metasurface dimensions are $P$ = 736 nm, $H$ = 695 nm and $L$ = 555 nm.

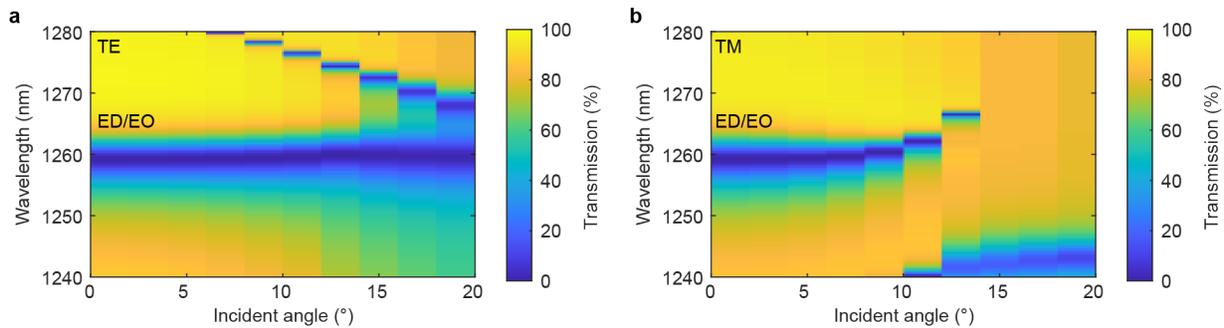

**Supplementary Figure 2 | Simulated angular dispersion for the metasurface in TE and TM polarization.** Simulated linear transmission ($T$) of the metasurface in Fig. 1b with $P$ = 736 nm, $H$ = 695 nm, $L$ = 555 nm for varying incident angles for TE (**a**) and TM (**b**) polarization.

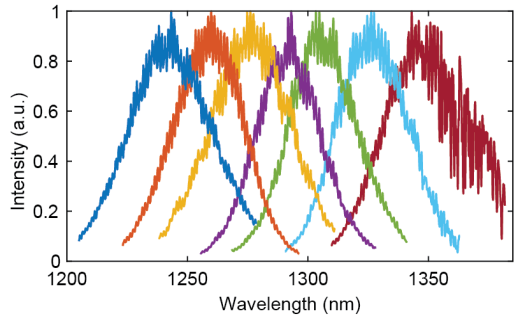

**Supplementary Figure 3 | NIR pump spectra.** Experimentally measured pump spectra corresponding to the resonant pump excitation of the metasurfaces illustrated in Fig. 2d.

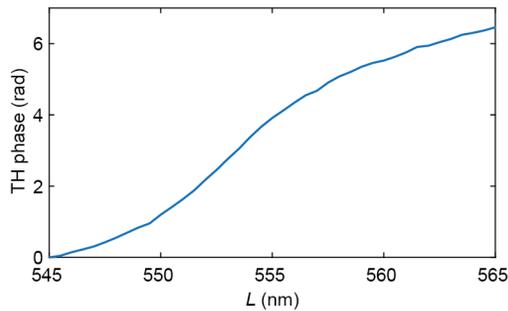

**Supplementary Figure 4 | TH phase look up table.** Simulated TH phase with varying nanoblock side length $L$, $P$ = 736 nm, and $H$ = 695 nm at a wavelength of $\lambda$ = 1260 nm, with the same geometrical parameters as in Fig. 1b. Note that the characterized the TH metalens in Fig. 4 is spectrally shifted with respect to this design wavelength to maximize THG due to critical coupling. Hence, we assume that the phase look up table does not significantly change with respect to the resonant wavelength.

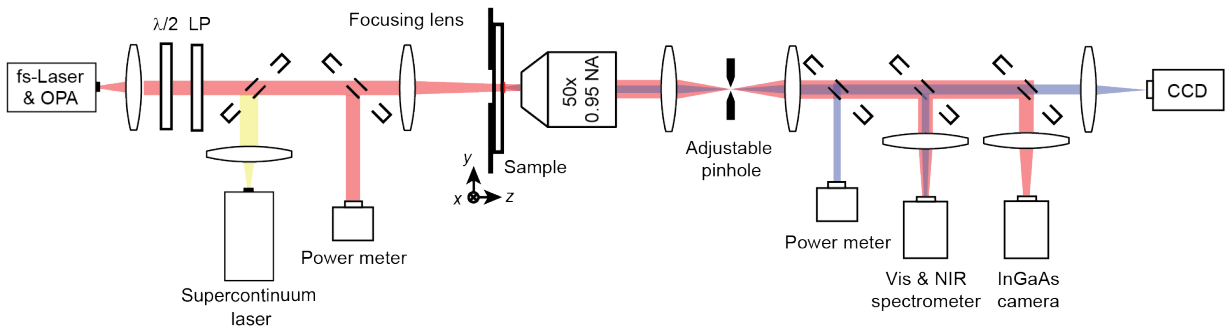

**Supplementary Figure 5 | Experimental set-up.** The fabricated samples are illuminated in transmission with loosely focused light from either a white light supercontinuum laser or a fs pulsed laser. The transmitted light is collected by an imaging objective (50x, 0.95 NA) and projected on the respective sensor for detection. With a set of flip mirrors, the transmitted light can be either sent to a visible camera, power meter or spectrometer, or then to a NIR camera or spectrometer. For the measurements presented in Fig. 3, the focusing lens is replaced with an objective lens (10x, 0.25 NA) and the numerical aperture of the illumination is set with an adjustable iris that is placed in front of the objective lens.

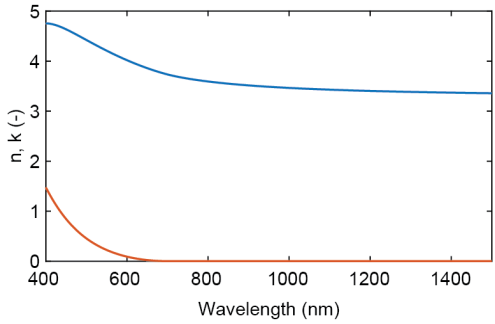

**Supplementary Figure 6 | Measured refractive index of amorphous silicon.** Real and imaginary refractive index of amorphous silicon as fitted to ellipsometry measurements using a Tauc-Lorentz model.

| | Method | TH Efficiency (×$10^{-6}$) | Peak Pump Intensity (GW/cm$^2$) | Wavefront manipulation |
|---|---|---|---|---|
| **Ref. 15** | Low order Mie resonances | Approx. ~1 | NA | Yes |
| **Ref. 23** | Guided mode | 0.176 | 0.52 | No |
| **Ref. 26** | EIT-analogue | 1.2 | 3.2 | No |
| **Ref. 35** | Higher order Mie resonances | 0.27 | 1.2 | No |
| **Ref. 16** | Low order Mie resonances | Approx. ~1 | 1.5 | Yes |
| **Ref. 29** | q-BIC | 1 | 0.1 | No |
| **Ref. 19** | Non-resonant geometric phase | $10^{-5}$ | 0.082 | Yes |
| **Ref. 31** | q-BIC | Approx. ~1 | 2.59 | No |
| **Ref. 18** | Low order Mie resonances | 1.1 | 33 | Yes |
| **Ref. 32** | q-BIC | 1.8 | 0.4 | No |
| **This work** | Higher order Mie resonances | 26.5 | 10.3 | Yes |

**Supplementary Table 1 | Comparison to the state of the art of THG with silicon based metasurfaces.** Comparison of the current state of the art of THG with silicon based metasurfaces. Experimentally reported values are compared.

| Lens diameter (μm) | Focal length (μm) | Numerical aperture | Fresnel Zones per surface | Nanoblocks per Fresnel Zone |
|---|---|---|---|---|
| 100 | 329.5 | 0.15 | 6 | 6–29 |

**Supplementary Table 2 | Metalens design parameters.** Design parameters for the metalens in Fig. 4. The periodicity of the nanoblocks is $P$ = 736 nm and $H$ = 695 nm. The phase distribution is parabolic and the nanoblock side lengths are set according to Supplementary Fig. 4.

## Supplementary Note 1: Numerical aperture dependent TH power

To provide further insight into the incident angle dependence of TH power from a metasurface we calculate the TH power emitted when illuminating with a beam with varying focusing numerical aperture. For this we assume a thin metasurface ($d$ << $\lambda$) and an illumination with a monochromatic gaussian beam of width $w$, an illumination cone angle $\theta$, and total incident power $P_\omega$. Additionally, we adopt the undepleted pump approximation. Furthermore, we adopt the hypothesis that the effective third order nonlinear susceptibility of the metasurface $\chi_{eff}^{(3)}$ is angle independent, which we then later test against the experimental data (see Fig. 3b). The electric field amplitude generated by the metasurface at the TH is approximated by[2,3]

$$E_{3\omega}(r) = \frac{i3\omega d}{8n_{3\omega}c}\chi_{eff}^{(3)}E_\omega^3(r), \quad (1)$$

where $E_\omega$ is the field from the gaussian illumination at the metasurface, $d$ is the thickness of the metasurface, and $n_{3\omega}$ is the refractive index at the TH. Equation (1) is used as an approximation as it assumes collimated illumination. The TH power, $P_{3\omega}$, is obtained by integrating the third harmonic intensity over the entire area of the illuminating beam

$$P_{3\omega} = \int \frac{1}{2}n_{3\omega}\varepsilon_0 c|E_{3\omega}(r)|^2 dA = \int \frac{9\varepsilon_0\omega^2 d^2}{128n_{3\omega}c}\left|\chi_{eff}^{(3)}E_\omega^3(r)\right|^2 dA, \quad (2)$$

where $n_\omega$ is the refractive index at the pump wavelength, $c$ represents the speed of light and $\varepsilon_0$ the vacuum permittivity. This can be rewritten to

$$P_{3\omega} = \int \frac{9\varepsilon_0\omega^2 d^2}{128n_{3\omega}c}\left|\chi_{eff}^{(3)}\right|^2 \left(\frac{2}{n_\omega\varepsilon_0 c}I(r)\right)^3 dA, \quad (3)$$

where $I(r) = \frac{2P_\omega}{\pi w^2}\exp\left(\frac{-2r^2}{w^2}\right)$ is the intensity of the gaussian beam. Inserted into the integral, the following expression is obtained

$$P_{3\omega} = \frac{9\omega^2 d^2}{16n_{3\omega}n_\omega^3\varepsilon_0^2 c^4}\left|\chi_{eff}^{(3)}\right|^2 \frac{8P_\omega^3}{\pi^3 w^6}\int_0^{2\pi}\int_0^\infty \left(e^{-\frac{2r^2}{w^2}}\right)^3 r\,dr\,d\theta. \quad (4)$$

This is then evaluated to

$$P_{3\omega} = \frac{3\omega^2 d^2 P_\omega^3}{4n_{3\omega}n_\omega^3\varepsilon_0^2 c^4\pi^2 w^4}\left|\chi_{eff}^{(3)}\right|^2 \sim NA^4. \quad (5)$$

## Supplementary Note 2: Effective third order nonlinear susceptibility

To determine the effective third order nonlinear susceptibility of the metasurface we follow a similar analysis as in the theoretical calculation described in Supplementary Note 1. The pump intensity $I(r)$ of the illuminating beam, the pump power $P_\omega$, and the TH power $P_{3\omega}$ are now known quantities from the measurement. However, since the camera only provides relative intensity data, the measured intensity of the camera, $I_{cam}$, needs to be renormalized to the total pump power by integration over the entire beam size

$$P_\omega = \int \frac{1}{2} n_\omega \varepsilon_0 c |E_\omega|^2 dA = \int \eta I_{cam} dA. \tag{6}$$

This allows determine the scaling factor $\eta$, which then allows obtaining the squared electric field amplitude

$$|E_\omega|^2 = \frac{2\eta}{n_\omega \varepsilon_0 c} I_{cam}. \tag{7}$$

With this adapted intensity we can rewrite Eq. (3) to

$$P_{3\omega} = \int \frac{9\varepsilon_0 \omega^2 d^2}{128 n_{3\omega} c} \left|\chi_{eff}^{(3)}\right|^2 \left(\frac{2\eta}{n_\omega \varepsilon_0 c} I_{cam}\right)^3 dA = \int \frac{9\omega^2 d^2 \eta^3}{16 n_{3\omega} n_\omega^3 \varepsilon_0^2 c^4} \left|\chi_{eff}^{(3)}\right|^2 I_{cam}^3 dA. \tag{8}$$

Solving this for $\chi_{eff}^{(3)}$ yields

$$\chi_{eff}^{(3)} = \sqrt{\frac{P_{3\omega}}{\int \frac{9\omega^2 d^2 \eta^3}{16 n_{3\omega} n_\omega^3 \varepsilon_0^2 c^4} I_{cam}^3 dA}}, \tag{9}$$

where the integral is evaluated numerically.

In the measurement, the pump power is measured before the focusing objective lens. As a result. We perform a separate measurement of the transmission of the focusing objective lens for varying numerical apertures to account for this effect when obtaining the pump power incident on the metasurface. The measured transmission of the focusing objective lens is illustrated in Supplementary Fig. 6.

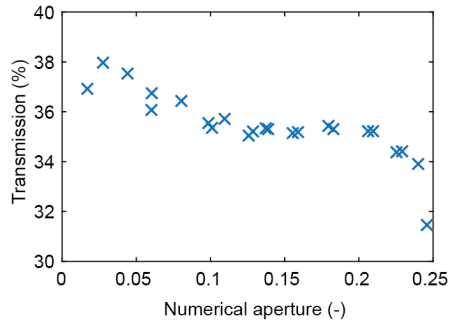

**Supplementary Figure 7 | Measured objective transmission vs NA at λ = 1327 nm.** For the measurement we use two identical colinearly aligned objective lenses (10x, 0.25 NA, Olympus) and measure the power before and after the two lenses for varying beam sizes, corresponding to the respective numerical aperture. This avoids errors induced by the incident angle dependent absorption usually encountered in photodiode power meters.

## References

1. Savinov, V., Fedotov, V. A. & Zheludev, N. I. Toroidal dipolar excitation and macroscopic electromagnetic properties of metamaterials. *Phys. Rev. B - Condens. Matter Mater. Phys.* **89**, (2014).
2. Boyd, R. W. *Nonlinear Optics*. *Academic Press* (Academic Press, 2008).
3. Kumar, N. *et al.* Third harmonic generation in graphene and few-layer graphite films. **121406**, 1–5 (2013).